\documentclass[sigconf]{acmart}
\AtBeginDocument{%
  }

\setcopyright{none}

\newcommand{\name}{CLIPGen}
\begin{document}

\title{\name{}: A \underline{C}hiplet \underline{L}ink \underline{IP} Modeling and \underline{Gen}eration Framework for 2.5D Architecture Exploration}


\author{Zhengping Zhu}
\email{zz2168@nyu.edu}
\affiliation{%
  \institution{New York University}
  \city{Brooklyn}
  \state{New York}
  \country{USA}
}

\author{Austin Rovinski}
\email{rovinski@nyu.edu}
\affiliation{%
  \institution{New York University}
  \city{Brooklyn}
  \state{New York}
  \country{USA}
}


\begin{abstract}
Advanced 2.5D Systems-in-Package (SiPs) compose a growing portion of high-performance systems.
While the packaging and interconnect choices play a large role in the overall system design, system architects still lack a suitable framework for early design space exploration which takes these choices into account.
Current interconnect models fall mostly into the categories of 1) detailed models which are generally inflexible and require deep packaging expertise, or 2) high-level models which don't provide enough information to make accurate architectural design decisions.

In this work, we present an automated chiplet IP generation framework which provides power, performance, and area estimates for various 2.5D packaging and communication configurations. The IP generator produces standard collaterals required for high-level simulation/estimation, RTL simulation, and place-and-route-level implementation (Verilog, Liberty, LEF, and datasheet). Using our framework, architects can co-optimize the package and chiplet architecture through rapid power, performance, and area estimates of various packaging strategies. As a case study, we examine generated UCIe interfaces across several packaging options.
\end{abstract}



\settopmatter{printacmref=false}
\renewcommand\footnotetextcopyrightpermission[1]{}
\fancyhead{}
\maketitle

\section{Introduction}

\begin{figure*}[!t]
    \centering
    \includegraphics[width=\linewidth]{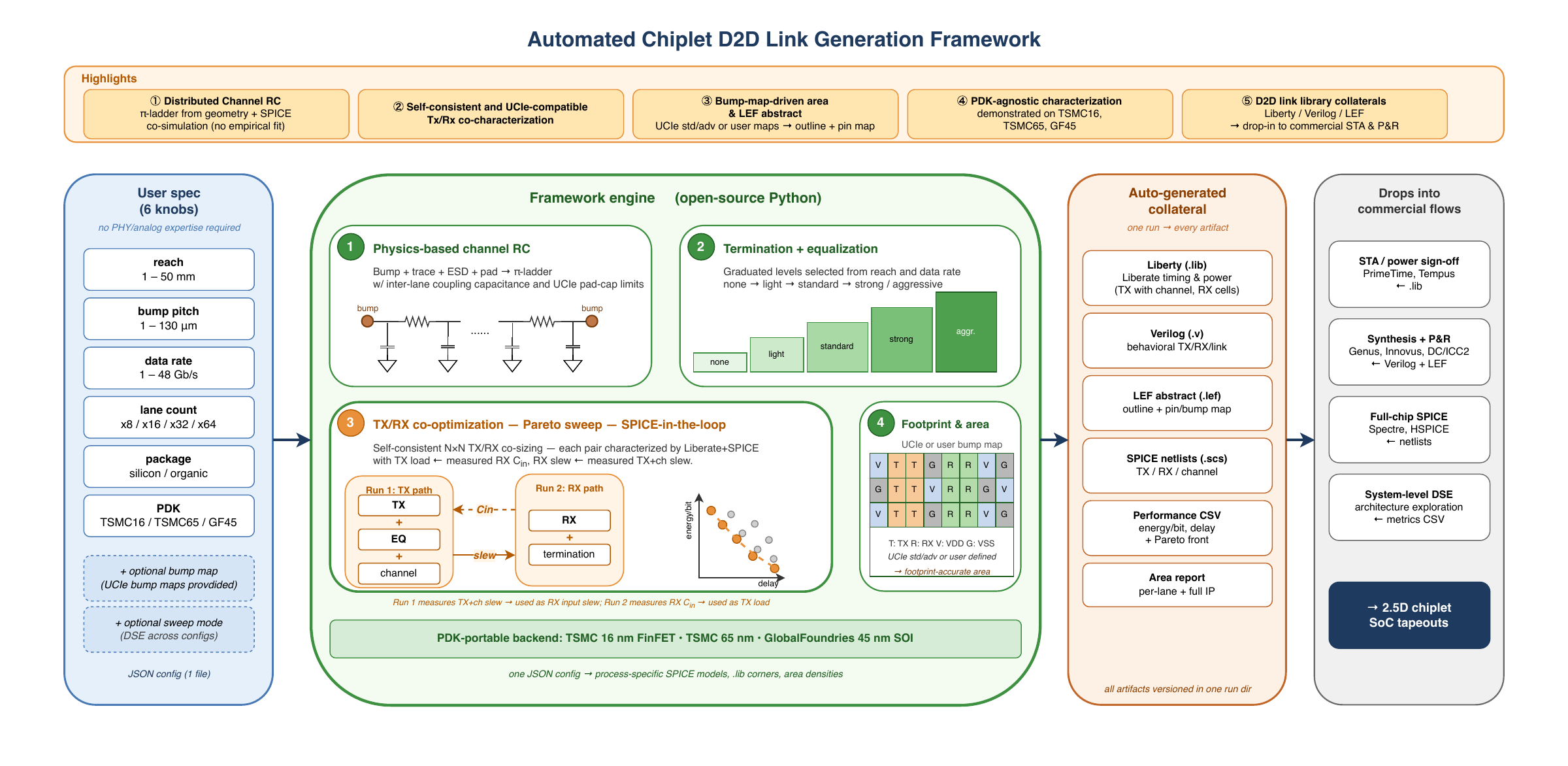}
    \caption{Overview of the proposed automated chiplet D2D link generation framework.
    A single JSON specification (reach, bump pitch, data rate, lane count, package type, PDK) drives:
    \textbf{(1)} a $\pi$-ladder channel RC derived from bump/trace/pad geometry and co-simulated in Liberate;
    \textbf{(2)} reach- and data-rate-driven auto-selection of termination and passive equalization;
    \textbf{(3)} \emph{self-consistent TX/RX co-characterization} over an $N{\times}N$ sizing grid, realized as two Liberate+SPICE runs that exchange measured RX $C_\text{in}$ and TX+channel slew, with a Pareto front filtered under the UCIe latency bound; and
    \textbf{(4)} bump-map-driven area estimation and LEF abstract generation (UCIe standard/advanced or user-supplied maps).
    The resulting \texttt{.lib}, Verilog, and LEF are drop-in for commercial STA and P\&R tools.
    The flow is PDK-agnostic and demonstrated on TSMC 16\,nm FinFET, TSMC 65\,nm, and GF 45\,nm SOI.}
    \label{fig:overview}
\end{figure*}

Chiplet-based systems have taken hold as a prime technology to enable continued scaling of computational systems beyond the end of Moore's Law. Chiplets offer several advantages over monolithic chips, including expanding logic beyond the die reticle limit, increasing yields above comparably sized monolithic systems, and increasing composability of systems by mixing/matching different chiplets during assembly.

However, system architects face significant challenges in performing design space exploration with different chiplet configurations to determine power, performance, and area (PPA) tradeoffs. Many prior works focus on either high-level modeling, which don't provide sufficient detail to make design decisions~\cite{Arch-DSEGemini-HPCA24,Arch-NNBaton,Arch-LUCIE-MICRO,Arch-DSEHISIM-TCAD}, or extremely deep modeling~\cite{zhang2018benchmarking}, which requires significant packaging expertise and effort from the architect.
Both ends of the spectrum often focus on a single technology stack, which significantly limits the design space when multiple packaging and process technologies are available at varying costs.

In this work, we present \name{}, a 2.5D chiplet I/O modeling framework and IP generator. The \name{} flow is shown in Fig. \ref{fig:overview}. \name{} takes in standard process design kit (PDK) files and a simple configuration file from the user as input. As output, it generates IP collaterals necessary for design space exploration at the pre-RTL, post-synthesis, and post-routing stages.
\name{} automatically generates datasheet (\texttt{.txt}), Verilog (\texttt{.v}), Liberty (\texttt{.lib}), physical abstract (\texttt{.lef}), and SPICE (\texttt{.scs}) formats. \name{} also generates a design constraint file (\texttt{.sdc}) which can be used to properly model interface constraints.

In addition, users with packaging and interconnect expertise can modify default model parameters (such as the degree of electrostatic discharge (ESD) protection) or even override internal parameters of the model to perform theoretical ``what if?'' exploration (such as the amount of coupling capacitance between channels).

This work bridges the gap not filled by existing frameworks -- we offer a multi-technology, multi-package I/O and link modeling framework which can offer accurate PPA metrics with negligible expertise from computer architects.
In this work, we offer the following contributions:

\begin{itemize}
    \item A chiplet link modeling framework which generates collaterals suitable for both early-stage and detailed PPA analysis, including datasheet, Verilog, Liberty, LEF, SPICE, and SDC
    \item A set of default, realistic packaging configurations to support system-level modeling and sweeps without requiring detailed packaging expertise.
    \item A detailed configuration framework which allows overriding modeling parameters and ``what if?'' exploration
    \item A case study of a UCIe interconnect across various packaging and technology configurations, demonstrating the importance of properly modeling interconnects at the architectural level.
\end{itemize}

\section{Related Works}


Prior modeling work in 2.5D chiplet I/O and link can be broadly classified into 3 categories:
1) Fabricated, lab-characterized chiplets,
2) Implementation-specific, detailed models, and
3) High-level modeling frameworks.


\subsection{Fabricated Chiplets}

Many prior works have explored chiplet interconnects and reported measurements for their systems~\cite{Design-Link-Simba,RW-IO-Jarr,Design-IO-VLSIGoyal,10181991,10904729}. Using these reported real measurements is enticing for architects, as it anchors their estimation in a ground truth. However, these works pose several issues for use in estimating PPA.
While there is a large body of work exploring more traditional 2D links (e.g. link length \textgreater{}40mm~\cite{Design-Link-Simba}), many fewer works report results in the 2.5D regime (link length \textless{}10mm~\cite{10181991,10904729}).
These works often report metrics piecemeal, with measurements such as I/O power remaining a single value and not breaking it down between components (TX, RX, link, protocol logic, etc.), or omit critical packaging details for contextualizing results.


Beyond missing data, characterized chiplets are also \textit{inflexible}, as they offer data for one very specific design point and don't offer insight for nearby design points. The data may also be subtly incompatible with the prototype system, due to a mismatch in process nodes (e.g. a 3nm IP on a 12nm chip) or other incompatibilties.

Our framework addresses this inflexibility by supporting arbitrary process nodes and a wide range of packaging options. All configuration options and metrics are reported to the user, and \name{} is calibrated against real works to present realistic data.

\subsection{Detailed Models}

An alternative approach to real-chiplet data is to use highly detailed models of chiplet interconnects.
Some works propose using schematic-based models which may include lumped or distributed parasitics models~\cite{Flow-Gatech-EPEPS,zhang2018benchmarking}.
While these works can offer high accuracy through detailed models, these models have cumbersome interfaces which require significant packaging expertise to understand. For example, a user may need to \textit{manually} supply bump, pad, and trace capacitances, and these values need to be accurate to receive reasonable results from the model. Architects are often not concerned with this level of detail, and may worry about receiving unrealistic results from a miscalibrated model.
Further, these models provide mainly the channel model and possibly a driver model. To the best of our knowledge, they do not model other commonly required high-speed I/O components such as ESD protection, equalizers, and termination, all of which impact the PPA of the chiplet link.

Our framework addresses this complication by providing simple, high-level parameters and reasonable default options for architects to use. The framework sources its internal properties from well-established physical constants (material permittivities, resistivity, trace parasitics per unit length), while still allowing users to override any parameters and provide their own values for calibration.

\subsection{High-level Models}

High-level modeling refers to models which use either static, fixed values, or very simple knobs to provide rapid estimates with minimal configuration.
High-level models are typically preferred by computer architects as they don't require expertise to configure or add to their system-level model. Many prior works simply use bandwidth numbers reported from prior works or specifications~\cite{ucie2.0-specification}. These works suffer from either lack of accuracy (especially relative to the I/O vs. logic), or lack of information, where the work may simply not account for or report important metrics like power and area~\cite{Arch-DSEGemini-HPCA24,Arch-NNBaton,Arch-LUCIE-MICRO,Arch-DSEHISIM-TCAD}.

Other times, works may select a single data point from a real chiplet system or detailed modeling framework~\cite{Arch-interface-Micro23}, which neglects any benefits of design space exploration. Further, the selected data point may differ in technology from other simulated results in the paper.

\subsection{Other Chiplet Models}
Several works have explored modeling other properties of 2.5D chiplets, such as cost and yield~\cite{feng2022chiplet,graening2023chiplets} and thermal properties~\cite{zhou2022thermal,pfromm2025mfit}. These works are orthogonal to this work, as they do not focus on the overall PPA modeling of chiplet interfaces. Many works also focus on 3D chiplet integration. While 3D integration is extremely promising for high performance interconnects, we focus on 2.5D modeling because 3D integration still faces steep challenges from thermal dissipation and yield. Additionally, interconnects in 3D systems generally are short enough (\textless{}150um) that they are firmly in the RC modeling regime and can be modeled fairly accurately as long on-chip wires.


\section{\name{} Framework}
\label{sec:framework}

\name{} follows four core design principles: (1)~\textbf{directional accuracy}—capture trends under parameter sweeps; (2)~\textbf{relative accuracy}—provide fair PPA comparison of I/O vs.\ logic; (3)~\textbf{ease of use}—provide realistic estimates without interconnect expertise or excessive runtime; (4)~\textbf{flexibility}—support multiple PDKs and parameter overrides.
Rather than aiming for tapeout-ready IP, \name{} captures realistic PPA into standard EDA collateral (Liberty, Verilog, LEF) suitable for system-level exploration with existing digital tools.

\name{} accepts a single JSON configuration file and produces a complete set
of standard EDA collateral---Liberty timing/power models, synthesizable
Verilog, LEF physical abstractions, and a metrics datasheet---for a
parameterized chiplet die-to-die (D2D) link IP.
The framework supports two operating modes:
(1)~\emph{single-point} mode, which generates and characterizes one link
configuration and optionally runs joint TX/RX co-optimization; and
(2)~\emph{sweep} mode, which evaluates a user-defined Cartesian product of
link parameters in parallel, producing a multi-dimensional design-space
dataset with a single invocation.

The end-to-end pipeline is illustrated in Fig.~\ref{fig:overview}.
Starting from a physical channel RC model, the framework proceeds through
automated link adaptation (termination and equalization), transceiver circuit
sizing, SPICE-level characterization via Cadence Liberate, and finally
collateral generation.
Each stage feeds the next: channel parasitics set the characterization load
for the TX Liberty run; the TX output slew in turn sets the RX
characterization index; and the resulting Liberty timing arcs compose the
full-link delay and energy budget reported in the datasheet.

\subsection{Configuration Interface}
\label{sec:config}

A deliberate two-tier parameter hierarchy separates \emph{user-facing}
parameters from \emph{hidden} physical constants.
User-facing parameters describe the target link in terms an architect
naturally controls: package type, reach, bump pitch, data rate, lane count,
and high-level design-intent flags (Table~\ref{tab:user_params}).
These are the only fields that must be set to explore a new configuration.

Hidden parameters (JSON sections suffixed \texttt{\_hidden}) encode
literature-sourced physical constants---dielectric permittivities, bump
resistivity, trace RC per unit length, and PHY boundary tables---that
are fixed for a given package technology.
Expert users may override them when calibrating the model to measured data,
but they need not be touched for ordinary design-space exploration.
This separation removes the requirement for packaging expertise to operate
the framework at the architect level.

\begin{table}[h]
\small
\caption{User-facing configuration parameters.}
\label{tab:user_params}
\begin{tabular}{lll}
\toprule
\textbf{Parameter} & \textbf{Type} & \textbf{Description} \\
\midrule
\texttt{pkg\_type}            & si $|$ org & Package / interposer technology \\
\texttt{reach\_mm}            & float               & Die-to-die interconnect length \\
\texttt{bump\_pitch\_um}      & float               & Bump-to-bump pitch \\
\texttt{data\_rate\_Gbps}     & float               & Per-lane NRZ data rate \\
\texttt{lane\_count}          & int                 & Number of D2D lanes \\
\texttt{passive\_eq\_en}      & bool                & Enable passive TX pre-emphasis \\
\texttt{ac\_coupled}          & bool                & Enable AC-coupled RX termination \\
\texttt{pad\_cap\_mode}       & phys $|$ ucie   & Geometry model or UCIe spec \\
\bottomrule
\end{tabular}
\end{table}

\subsection{Operating Modes and Options}
\label{sec:modes}

Beyond the eight core link parameters, \name{} exposes a rich set of
operating modes that control channel modeling fidelity, transceiver sizing
strategy, characterization accuracy, physical layout, and output artifacts.
All modes have reasonable defaults so that a minimal config runs end-to-end
without modification; users selectively enable or tune options as their
exploration matures.

\textbf{Channel fidelity options.}
The pad capacitance model is switchable between a geometry-derived
parallel-plate model (\texttt{pad\_cap\_\allowbreak{}mode: physical}) and a UCIe
Standard Package specification lookup table (\texttt{pad\_cap\_mode: ucie}),
which returns the spec-mandated maximum pad cap (300/200/125\,fF) for the
configured data rate.
The \texttt{ucie} mode is useful for compliance-centric analysis where the
pad cap budget is fixed by the standard rather than the physical geometry;
it also subsumes ESD capacitance, which the physical model tracks separately.

Inter-lane coupling capacitance can be enabled or disabled as a block
(\texttt{coupling\_cap.enabled}).
When enabled, three knobs are exposed: \texttt{cc\_ratio\_trace}
(coupling fraction at each trace Pi-ladder node, typically 0.3--0.5 for
edge-coupled microstrip), \texttt{cc\_ratio\_pad} (fraction at interposer
pad nodes, smaller since pads are more widely spaced than traces), and
\texttt{cc\_rx\_pad\_fF} (absolute on-die RX pad-to-pad fringe capacitance
in fF).
The coupling elements are injected directly into the SPICE netlist so that
Liberate captures their effect on TX output transition time and RX input
sensitivity.

Geometric channel parameters---bump diameter, bump height, trace width,
and ESD type---all default to package-derived values but may be overridden
individually when the physical design deviates from the default rules.

\textbf{Transceiver sizing hierarchy.}
\name{} offers four progressively more powerful sizing strategies,
summarized in Table~\ref{tab:sizing_modes}.
Users start with the simplest mode that meets their needs and escalate
as the design matures.

\begin{table}[h]
\small
\caption{Transceiver sizing modes, in order of increasing automation.}
\label{tab:sizing_modes}
\begin{tabular}{p{1.6cm}p{3.5cm}p{2.5cm}}
\toprule
\textbf{Mode} & \textbf{Description} & \textbf{Best for} \\
\midrule
Manual &
  User specifies TX and RX transistor widths directly &
  Porting a known design \\
\texttt{tx\_sizing} &
  Adaptive SPICE search (golden-section) to meet UCIe rise/fall target
  (\texttt{rise\_fall\_pct\_ui}) within a simulation budget &
  TX-only tuning \\
\texttt{rx\_sizing} &
  Iterative RX width search to satisfy a maximum RX delay fraction of UI &
  RX-only tuning \\
\texttt{co\_opt} &
  Joint TX+RX Pareto optimization via LUT matching;
  supersedes the two modes above &
  Full link optimization \\
\bottomrule
\end{tabular}
\end{table}

\noindent
The \texttt{co\_opt} mode additionally exposes \texttt{n\_tx\_configs} and
\texttt{n\_rx\_\allowbreak{}configs} (number of configurations in the exploration grid),
\texttt{max\_\allowbreak{}parallel} (concurrent characterization processes), and
\texttt{pareto\_\allowbreak{}selection} (\texttt{balanced}, \texttt{best\_power},
\texttt{best\_delay}, or \texttt{all} to return the full frontier).

\textbf{Characterization options.}
The standard mode embeds the channel RC $\pi$-ladder inside the TX SPICE subcircuit, so Liberate characterizes the full TX$+$channel path in one run, producing a single timing arc that captures channel loading inherently.
(A legacy \texttt{false} option exists for external analytical correction, but is non-standard.)
The \texttt{rx\_slew\_source} option selects whether RX input slews are drawn from the TX pad before the channel (\texttt{tx\_pad}) or from the far end of the $\pi$-ladder (\texttt{channel}); the latter is physically correct.
An explicit \texttt{input\_slews\_ns\_override} list may override auto-detection if needed.

\textbf{Physical layout mode.}
When \texttt{bump\_map\_enabled} is set, \name{} reads a user-supplied bump
map text file that assigns each bump site to a role: \texttt{tx},
\texttt{rx}, \texttt{vdd}, \texttt{vss}, \texttt{other}, or empty.
The bump count for \texttt{tx} and \texttt{rx} roles must match
\texttt{lane\_count}; power and ground bump counts are free.
The area model and LEF generator use the bump map to produce a
full-footprint macro bounding box with correct pin placement for each
role, enabling accurate floorplan integration.
Without a bump map, only per-lane area is reported.

\textbf{Output control.}
Each output artifact is independently toggled: SPICE netlists
(\texttt{save\_netlists}), Liberate characterization scripts
(\texttt{save\_liberate\_decks}), Liberty files (\texttt{save\_lib}),
aggregate metrics CSV (\texttt{save\_metrics\_csv}), Verilog
(\texttt{generate\_verilog}), and LEF (\texttt{generate\_lef}).
This allows, for example, a sweep run to save only the CSV for rapid
analysis while suppressing the larger binary and netlist artifacts.

\subsection{Physical Channel Model}
\label{sec:channel}

\begin{figure}
    \centering
    \includegraphics[width=\linewidth]{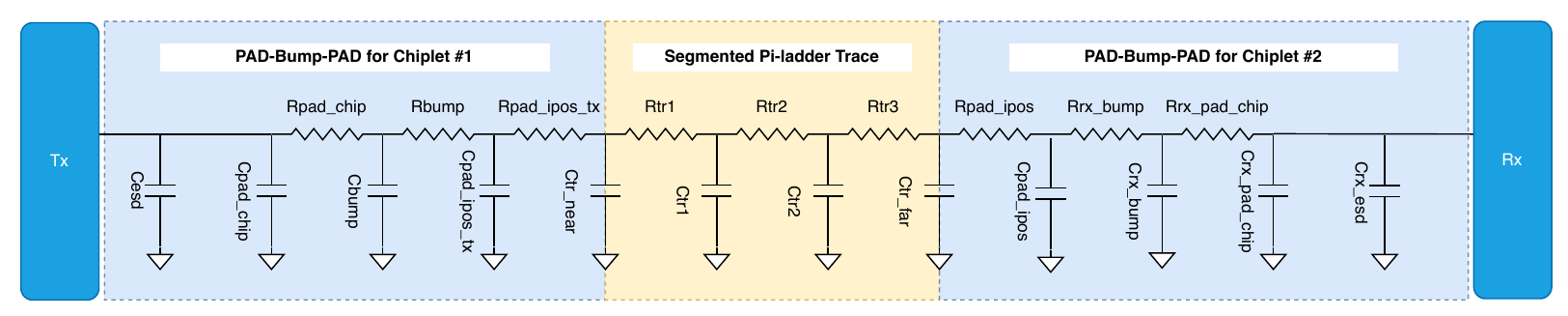}
    \caption{Distributed $\pi$-ladder channel model. Per-element R$+$C for chiplet and interposer pads, microbumps, and ESD shunt caps at each chiplet side; three-segment $\pi$-ladder for the trace (near half-cell, three series resistors with shunt caps, far half-cell).}
    \label{fig:channel_model}
\end{figure}

Full-wave electromagnetic (EM) tools can model 2.5D interconnects with high accuracy but require hours per configuration, unsuitable for design-space exploration.
For the UCIe Standard regime ($\leq$32\,Gbps, $\leq$50\,mm), the electrical length is well under one-tenth the signal wavelength at Nyquist; RC parasitics dominate the signal integrity mechanism, with sufficient accuracy for this design space~\cite{Dehlaghi2019USR}.
Inductive effects are negligible; skin-effect-corrected bump resistance is used.

The channel model computes per-segment RC values for each 2.5D component: chiplet pad, microbump, interposer pad, trace (three segments), and ESD protection.
These values are assembled into a distributed $\pi$-ladder SPICE subcircuit (\texttt{txip.scs}) and embedded directly in the TX characterization netlist, so Liberate produces timing arcs that include channel propagation without external analytical correction.
The same per-segment RC values also support rapid Elmore-based pre-filtering during design-space exploration.
All components are parameterized by user-facing geometry (bump pitch, reach) together with package-type-dependent constants from the hidden config.

\textbf{Per-component models.}
All physical constants (material permittivities, resistivity, trace parasitics, interposer layer thicknesses, ESD targets) are sourced from literature~\cite{IBM2006Wright,kim2011high,tummala2001fundamentals,ESDLeuven,Dehlaghi2019USR}, ensuring calibration against industry-standard references and enabling multi-technology portability.

Pad capacitance is modeled as a parallel-plate capacitor scaled by pad width
$w_p = 0.8 \cdot P$ (bump pitch $P$) and ILD thickness $t_{ox}$:
\begin{equation}
C_{\mathrm{pad}} = \varepsilon_0 \varepsilon_{r} \frac{w_p^2}{t_{ox}}.
\label{eq:cpad}
\end{equation}
Pad resistance scales inversely with $w_p$ from a reference measurement.
Bump capacitance uses a two-wire cylindrical model with underfill permittivity;
bump resistance accounts for both DC and skin-effect (AC) contributions at
the Nyquist frequency $f_{Ny} = R_b/2$:
\begin{equation}
R_{\mathrm{bump}} = \sqrt{R_{DC}^2 + R_{AC}^2}.
\label{eq:rbump}
\end{equation}
Trace capacitance and resistance are scaled from tabulated base values
(silicon interposer: 185\,fF/mm, 1.04\,$\Omega$/mm at 3\,$\mu$m width;
organic substrate: 138\,fF/mm, 0.036\,$\Omega$/mm at 30\,$\mu$m
width~\cite{Dehlaghi2019USR}) by width and dielectric ratio.
ESD capacitance is selected from a graduated table keyed on bump pitch and
package type, reflecting CDM target-level specifications~\cite{ESDLeuven}.

\textbf{SPICE simulation.}
The per-segment RC values are assembled into a three-segment $\pi$-ladder SPICE subcircuit embedded in \texttt{txip.scs}, which Liberate co-characterizes with the TX driver at each configuration point. This produces Liberty timing arcs that intrinsically capture the channel's effect on TX slew and delay, with no external analytical correction needed.

\textbf{Analytical filtering (optional).}
For rapid Pareto pre-filtering during design-space sweeps, a lumped Elmore reduction over the same RC values provides sub-millisecond estimates of delay and energy:
\begin{align}
\tau_{\mathrm{Elmore}} &= R_{ch} \cdot C_{ch}, \label{eq:elmore} \\
E_{\mathrm{ch}} &= \tfrac{1}{2} \, C_{\mathrm{eff}} \, V_{DD}^2, \label{eq:ech}
\end{align}
where $C_{\mathrm{eff}}$ counts the chiplet pad, microbump, and interposer pad twice each (charged and discharged per bit) and single-counts the trace and ESD.
This analytical estimate pre-filters the design space; final timing and energy metrics come from the full SPICE $\pi$-ladder via Liberty.

\textbf{Configuration options.}
An alternative \emph{UCIe pad cap mode} replaces the geometry-derived pad capacitance with the UCIe Standard Package specification lookup table (300\,fF at $\leq$8\,GT/s; 200\,fF at $\leq$16\,GT/s; 125\,fF at $\leq$32\,GT/s), absorbing ESD into the spec budget.
Inter-lane coupling capacitance can be enabled as a ratio of per-node shunt capacitance, injected into the SPICE $\pi$-ladder as coupling elements between adjacent lanes.

\subsection{Automated Link Adaptation}
\label{sec:adaptation}

Given the computed channel RC, \name{} automatically determines whether termination and equalization are needed and sizes their components from first-principles models.
The user only sets high-level flags (\texttt{passive\_eq\_enabled}, \texttt{ac\_coupled}); all thresholds and values are computed by the framework.

\textbf{Termination.}
For short-reach links within the UCIe unterminated operating boundary, impedance matching is unnecessary and termination adds power overhead without SI benefit.
Beyond the boundary, reflection suppression is required.
The Thevenin-split topology (two high-impedance bias resistors to mid-rail, plus termination resistor $R_T$) is UCIe-specified and avoids DC bias current, keeping static power negligible~\cite{ucie2.0-specification}.
Rather than binary on/off, \name{} uses a graduated four-level scheme: as reach exceeds the unterminated boundary, stronger termination (lower $R_T$) is selected progressively, matching overhead to actual SI need.
The need and level are decided by a UCIe-based boundary table indexed by TX swing and data rate, yielding the maximum unterminated reach
$L_{\mathrm{unterm}}$. The reach ratio $\rho = L / L_{\mathrm{unterm}}$ selects one of four graduated levels:

\begin{table}[h]
\small
\caption{Graduated termination levels.}
\label{tab:term_levels}
\begin{tabular}{llll}
\toprule
\textbf{Level} & \textbf{$\rho$ range} & \textbf{$R_{\mathrm{term}}$} & \textbf{$C_{AC}$} \\
\midrule
0 -- none     & $\leq 1.0$  & ---          & ---            \\
1 -- light    & $\leq 1.25$ & $2 R_{RX}$   & $0.5 C_{base}$ \\
2 -- standard & $\leq 1.5$  & $R_{RX}$     & $C_{base}$     \\
3 -- strong   & $> 1.5$     & $0.5 R_{RX}$ & $2 C_{base}$   \\
\bottomrule
\end{tabular}
\end{table}

\noindent
The Thevenin-split topology sets $V_{TERM} = V_{DD}/2$ via
high-im\-pedance bias resistors (1\,M$\Omega$), keeping static bias
current negligible.
Termination energy per bit combines the static bias dissipation and the
dynamic switching component through $R_{\mathrm{term}}$.

\textbf{Equalization.}
Active equalization (FFE) requires a DAC, multi-tap serializer, and significant power/area overhead.
For short-reach, moderate-loss channels (dominant degradation: first-order RC roll-off, not multi-pole), a single-zero passive equalizer suffices: a series poly-R and shunt MIM/MOM-C at the TX output create a high-pass response that cancels the RC roll-off with only two passive components.
Like termination, equalizer strength is graduated (five levels: none through aggressive) and auto-selected based on channel loss.
Short, low-loss links engage no EQ; longer/higher-rate links engage progressively stronger pre-emphasis, avoiding unnecessary area on links that don't need it.

The channel RC low-pass loss at Nyquist,
\begin{equation}
\mathrm{Loss}_{Ny} = 10\log_{10}\!\left[1 + \left(\frac{f_{Ny}}{f_{3dB}}\right)^{\!2}\right],
\label{eq:loss}
\end{equation}
is compared against a configurable threshold to select one of five graduated
passive EQ levels (\emph{none} through \emph{aggressive}).
The selected level maps to an equalizer capacitance fraction $\alpha$ of $C_{\mathrm{ch}}$,
where
\[
  \alpha \in \{0,\;0.05,\;0.10,\;0.15,\;0.20\}.
\]
Component values are derived by placing the EQ zero at the channel $-3$\,dB
corner:
\begin{equation}
C_{eq} = \alpha \, C_{ch}, \quad
R_{eq} = \frac{R_{ch}}{\alpha}.
\label{eq:eq_sizing}
\end{equation}
$R_{eq}$ is additionally capped so that $R_{eq} \cdot C_{\mathrm{downstream}}$
does not exceed a latency budget multiple of the unit interval,
preventing excessive pre-emphasis on long-reach channels.

\subsection{Transceiver Sizing and Characterization}
\label{sec:xcvr}

\textbf{TX Architecture.}
For short-reach D2D links at UCIe Standard data rates and bump pitches, pad capacitance is modest (125--300\,fF) and signal swing is full-rail CMOS.
A tapered CMOS inverter chain is the dominant TX architecture in published designs~\cite{ucie2.0-specification}: lowest static power (no DC bias), simplest analytical sizing, direct Liberty compatibility.
(High-swing differential topologies offer better SI at longer reaches but incur static power and different termination strategy; they are outside the current Standard Package scope.)

The chain fanout factor is sized analytically following geometric progression $f \approx e \approx 2.718$ (adjusted for parasitic capacitance), with stage count forced even to maintain non-inverting polarity at the pad.
A SPICE Q/V simulation measures the unit inverter input capacitance; the full chain is then sized geometrically.
When passive EQ is enabled, the computed $R_{eq}$ and $C_{eq}$ are patched into the TX netlist template.
The channel RC $\pi$-ladder is embedded directly in the TX subcircuit so that a single Liberate run characterizes the full TX$+$channel path.

\textbf{RX Architecture.}
The RX faces two competing constraints: minimize input capacitance seen by the channel (preserve bandwidth), yet drive core logic fanout.
A single inverter cannot do both; a two-stage topology resolves this: small pre-amplifier (minimal input cap) + larger output buffer (core fanout).
This standard short-reach RX architecture~\cite{ucie2.0-specification,Dehlaghi2019USR} maps naturally to a two-cell Liberty abstraction with separate input/output characterization points.

Input slew for RX characterization is auto-detected from the TX output waveform at the far end of the channel $\pi$-ladder, ensuring consistent characterization conditions without manual input specification.
An optional sizing mode iterates buffer widths to satisfy a maximum RX delay fraction of the unit interval.

\textbf{Area model.}
The silicon footprint of each TX and RX IP macro is computed analytically
from the sized component values, using process-specific density constants
from the \texttt{area\_hidden} config section:
active transistor area (with a layout margin factor for contacts, poly
extensions, and well taps), passive EQ (poly-R strip and MIM/MOM capacitor),
termination bias resistors, and ESD diode area derived from pad capacitance.
Bump and UBM pad geometry follows assembly-regime rules: Cu pillar/C4 for
pitch $P \geq 10$\,$\mu$m, hybrid bonding (Cu--Cu direct bond, no UBM
overhang) for $P < 10$\,$\mu$m.
The bounding box dimensions feed directly into the LEF macro generator.

\subsection{TX/RX Co-Optimization via Pareto Search}
\label{sec:coopt}

TX and RX sizing are mutually coupled: TX delay and output slew depend on
the RX input capacitance (output load), while RX delay depends on the TX
output slew (input transition time).
Sizing each independently ignores this coupling and produces suboptimal
results.
\name{} formulates the joint optimization as a multi-objective problem over
the Pareto frontier of total energy per bit and worst-case link latency.

\textbf{Lookup-table matching.}
Rather than running a Liberate characterization for every (TX$_i$, RX$_j$)
pair, \name{} exploits the separability of the characterization tables.
Each TX configuration is characterized \emph{once} with a sweep of output
loads covering the full RX input capacitance range; each RX configuration
is characterized \emph{once} with a sweep of input slews covering the full
TX output slew range.
For each pair, metrics are obtained in sub-millisecond time by
interpolation:
\begin{enumerate}
  \item Read RX$_j$ input capacitance $C_{RX}$ from its Liberty attribute.
  \item Interpolate TX$_i$ tables at load $= C_{RX}$ $\rightarrow$ TX delay,
        output slew, switching power.
  \item Interpolate RX$_j$ tables at input slew $=$ TX$_i$ output slew
        $\rightarrow$ RX delay, switching power.
  \item Sum delay; compute total $E/\mathrm{bit} = \alpha (E_{TX} + E_{RX}) +
        E_{ch} + E_{term}$.
\end{enumerate}
Total Liberate invocations: $N_{TX} + N_{RX}$, versus
$\mathcal{O}(N_{TX} \times N_{RX})$ for a brute-force grid.

\textbf{Pareto frontier.}
The feasible set (configurations satisfying the UCIe latency constraint)
is reduced to a non-dominated frontier on
$\bigl(E_{\mathrm{total}}\,[\mathrm{pJ/bit}],\ \tau_{\mathrm{wc}}\,[\mathrm{ps}]\bigr)$,
where $\tau_{\mathrm{wc}} = \max(\tau_{RR}, \tau_{FF})$ is the worst-case
TX$+$channel$+$RX delay.
Three automated selection strategies are provided: \texttt{balanced}
(minimum Euclidean distance to the ideal point), \texttt{best\_power},
and \texttt{best\_delay}.

\subsection{Multi-PDK Portability}
\label{sec:pdk}

\name{} provides production-quality configurations for three process nodes:
TSMC~65\,nm planar CMOS, TSMC~16\,nm FinFET, and GF~45\,nm SOI.
Process-specific differences are fully encapsulated within each config file
and the corresponding Liberate characterization template (SPICE netlist,
characterization script):
FinFET fin-count quantization (e.g., $w = 0.01 + 0.048(n_{fin}-1)$\,$\mu$m),
MIM vs.\ MOM capacitor density, BEOL ILD thickness, and supply voltage
scaling are all handled automatically.
The analytical models (channel, equalization, termination, area) are
process-agnostic and require no modification when switching nodes.

Porting to a new process node requires authoring a JSON config and a
Liberate netlist template.
The JSON config supplies transistor names, minimum/maximum widths,
PDK model file path, and process corner; the template provides a unit
TX inverter subcircuit that the framework wraps into the multi-lane
\texttt{txip} netlist and patches with the computed EQ and termination
parameters.
\subsection{Design Space Exploration and Collateral Generation}
\label{sec:dse}

In sweep mode, \name{} evaluates the full Cartesian product of
$\{\texttt{pkg\_type}\} \times \{\texttt{reach\_mm}\} \times
\{\texttt{bump\_pitch\_um}\} \times \{\texttt{data\_rate\_Gbps}\}$
in parallel, limited by a configurable worker count.
Each point independently traverses the full pipeline---channel model,
adaptation, characterization, co-optimization---and writes its results
to a per-configuration subdirectory.
An aggregate CSV collects all metrics, enabling cross-sweep Pareto
analysis and visualization (e.g., energy vs.\ reach across package
technologies, EQ/termination regime maps).

For every evaluated configuration, \name{} produces the following
standard collateral:
\begin{itemize}
  \item \textbf{Liberty (.lib):} NLDM timing and internal switching power
        for \texttt{txip} and \texttt{rxip}, compatible with Synopsys
        PrimeTime and open-source OpenSTA.
  \item \textbf{Verilog:} Behavioral RTL model with correct port naming
        and bus widths for \texttt{lane\_count} lanes, suitable for
        RTL-level simulation.
  \item \textbf{LEF:} Physical macro abstraction with pin locations derived
        from the bump map and area model bounding box, suitable for
        place-and-route floorplanning.
  \item \textbf{Datasheet / CSV:} Per-configuration PPA summary including
        per-lane and aggregate energy breakdown, latency, area, and
        the engaged termination and equalization levels.
\end{itemize}

\begin{figure}[t]
    \centering
    \includegraphics[width=\columnwidth]{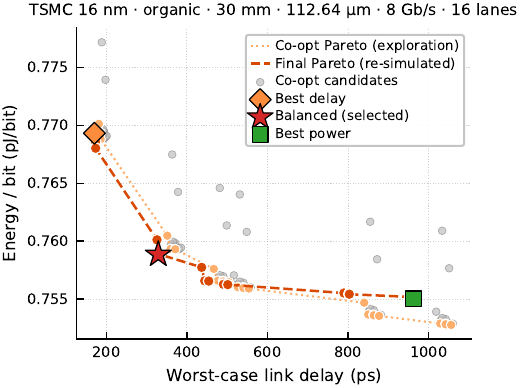}
    \caption{%
        Energy-per-bit vs.\ worst-case link delay for a 16-lane UCIe Standard
        link (TSMC 16\,nm, organic substrate, 112.64\,\textmu{}m bump pitch,
        8\,Gb/s, 30\,mm reach). Grey dots are all TX/RX sizing
        candidates examined during co-optimization. The light-orange dotted
        curve is the Pareto frontier estimated during the exploration phase;
        the dark-orange dashed curve is the frontier after the final
        re-characterization with matched TX output slew and RX input capacitance.
        Three representative operating points are highlighted:
        minimum-delay (orange diamond), balanced energy--delay knee
        (red star, auto-selected), and minimum-energy (green square).
    }
    \label{fig:pareto_frontier}
\end{figure}
\section{Experiments and Results}
\label{sec:experiment}

\subsection{Pareto-Guided Design-Point Selection}
\label{sec:pareto_selection}

Figure~\ref{fig:pareto_frontier} shows how \name{} converts a single link
configuration into a family of Pareto-optimal design points. For the
30\,mm organic UCIe Standard link at 8\,Gb/s, the $N{\times}N$
co-optimization grid (Section~\ref{sec:coopt}) produces dozens of
feasible TX/RX pairs (grey cloud) spanning more than $2{\times}$ in
worst-case delay. The non-dominated subset forms the Pareto frontier;
\name{} exposes three canonical selections (minimum delay, minimum
energy, balanced knee) and emits the full frontier as collateral.

The two frontiers in Fig.~\ref{fig:pareto_frontier} differ slightly
because exploration approximates the per-stage TX load and RX input
slew from shared lookup tables to keep the $N{\times}N$ sweep
tractable, while the re-characterization step re-simulates each
surviving pair with matched $C_\text{in}$ and TX+channel slew. The
shift is typically a few percent in energy and tens of picoseconds in
delay, and the final curve is what populates the generated
\texttt{.lib}, \texttt{.lef}, and Verilog collateral.

\begin{figure}[t]
    \centering
    \includegraphics[width=\columnwidth]{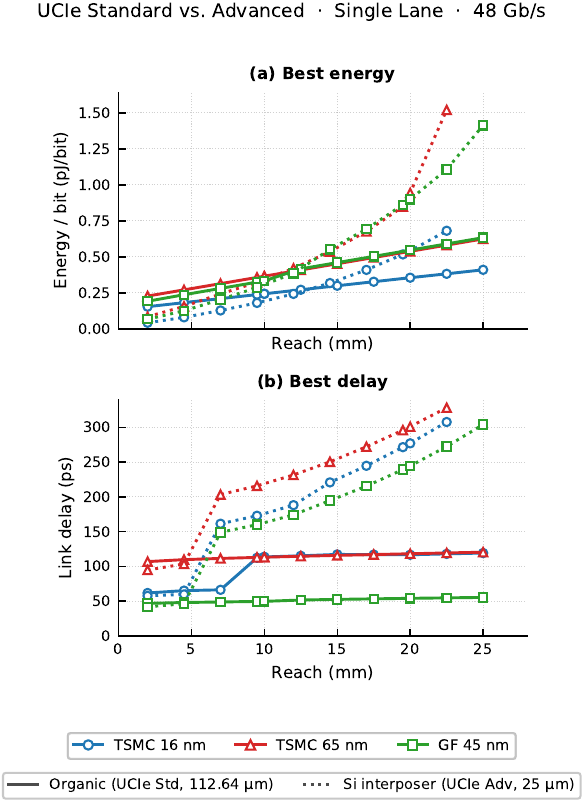}
    \caption{%
        Single-lane, 48\,Gb/s short-reach comparison of an organic UCIe
        Standard substrate (112.64\,\textmu{}m bump pitch, solid lines)
        and a silicon interposer (25\,\textmu{}m bump pitch, dotted
        lines) across three PDKs.
        (a) Energy per bit of the \emph{best-power} Pareto selection.
        (b) Worst-case link delay of the \emph{best-delay} Pareto
        selection. Only delay-feasible points under the UCIe 16\,UI
        latency budget are plotted.
    }
    \label{fig:short_reach_pkg}
\end{figure}
\subsection{Short-Reach: When Does an Organic Substrate Beat a Silicon Interposer?}
\label{sec:short_reach_pkg}
Conventional wisdom holds that a silicon interposer (UCIe Advanced,
25\,\textmu{}m pitch) is always superior to an organic substrate
(UCIe Standard, 112.64\,\textmu{}m pitch) at short reach because of its
lower bump capacitance. Figure~\ref{fig:short_reach_pkg} tests this on
a single-lane 48\,Gb/s link sweeping reach from 2 to 25\,mm on both
packages across the same three nodes, plotting only UCIe-feasible
points.

Contrary to intuition, the two curves cross: for all three PDKs the
best-power energy curves intersect near \textbf{10\,mm}, and at 25\,mm
the gap exceeds 2$\times$ in the organic's favour (e.g.\ 0.41 vs.\
0.83\,pJ/bit on TSMC 16\,nm; 0.63 vs.\ 1.76\,pJ/bit on TSMC 65\,nm).
Silicon has a small delay edge only at 2--4.5\,mm, after which its
best-delay jumps to $\sim$150--200\,ps while organic stays in the
50--120\,ps range out to 25\,mm.

The modeling flow (Section~\ref{sec:channel}) explains the crossover
via two opposing terms. \emph{Bump/pad capacitance}
(Eqs.~\eqref{eq:cpad}--\eqref{eq:rbump}): the 112.64\,\textmu{}m organic
bump costs $\approx 63$\,fJ/bit at the pad versus $\approx 4.5$\,fJ/bit
for the 25\,\textmu{}m silicon bump---a fixed $\sim$60\,fJ/bit penalty
organic always pays. \emph{Distributed channel RC}
(Section~\ref{sec:channel}): the fine-pitch silicon wires have
much higher resistance per mm, so the Elmore
constant~\eqref{eq:elmore} degrades rapidly and the co-optimizer must
drive the channel harder. On TSMC 16\,nm best-power, silicon
$E_{\mathrm{TX}}$ rises from 0.8\,fJ/bit (2\,mm) to 28\,fJ/bit
(9.5\,mm), 234\,fJ/bit (20\,mm), and over 2\,pJ/bit by 42.5\,mm; organic
stays in single-digit fJ/bit until $\sim$30\,mm. At $\sim$10\,mm the
silicon TX blow-up has erased its bump-cap advantage, and for longer
reach the thicker organic trace's lower resistance dominates. This is
exactly the package-level trade-off that is tedious to hand-calibrate
but falls out naturally from the parameterized channel and TX/RX
co-optimization in \name{}.

\begin{figure*}[t]
    \centering
    \includegraphics[width=\textwidth]{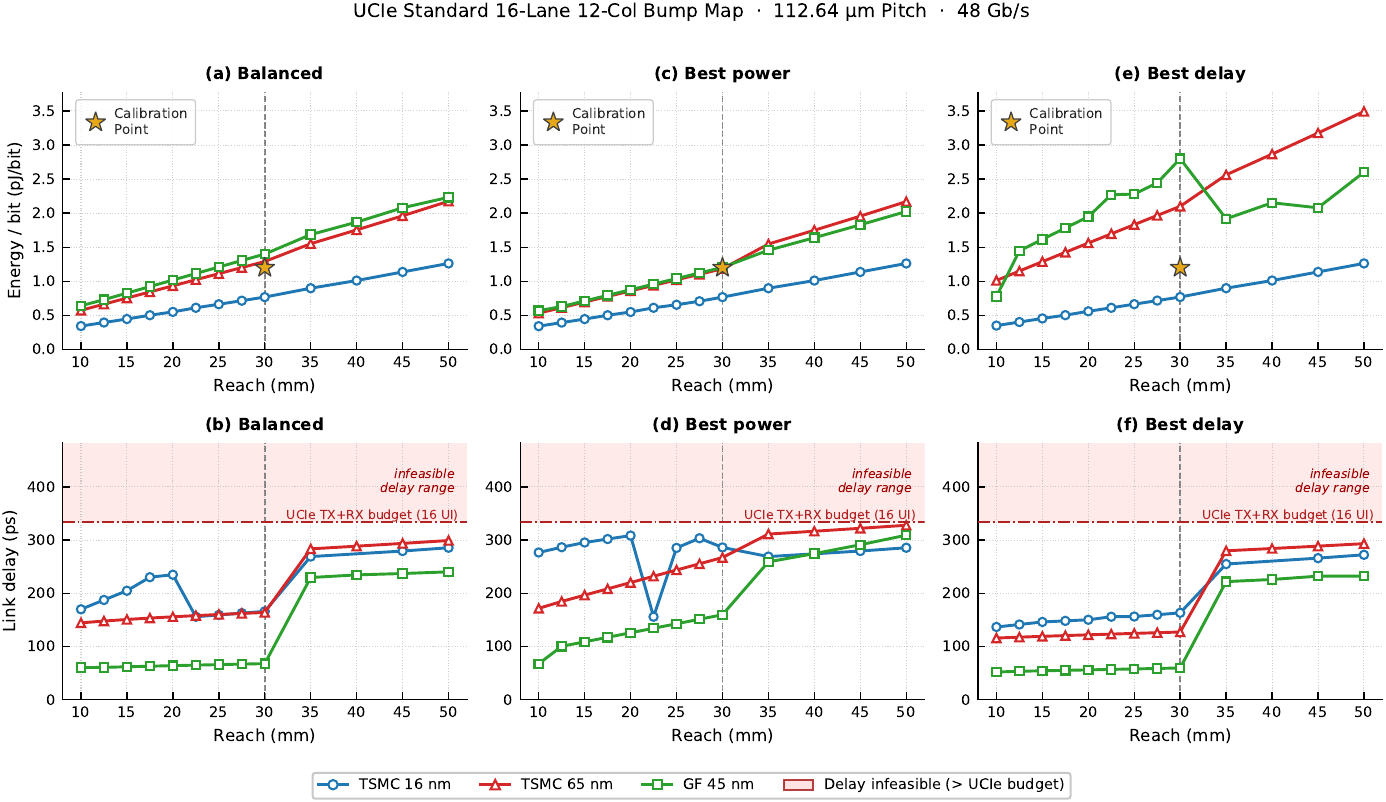}
    \caption{%
        Modeled link energy per bit (top row) and worst-case link delay (bottom row)
        versus reach for a 16-lane chiplet link at 48\,Gb/s on an organic
        UCIe Standard substrate (12-column x16 bump map, 112.64\,\textmu{}m bump
        pitch), swept across three process nodes: TSMC 16\,nm (VDD\,$=$\,0.8\,V),
        TSMC 65\,nm (VDD\,$=$\,1.0\,V), and GF 45\,nm (VDD\,$=$\,1.0\,V).
        Columns correspond to three Pareto-optimal design-point selections:
        (a,\,d)~balanced energy--delay trade-off,
        (b,\,e)~minimum-energy (\emph{best-power}), and
        (c,\,f)~minimum-delay (\emph{best-delay}).
        The red dash-dot line in the delay panels marks the UCIe TX+RX latency
        budget of 16\,UI ($\approx$333\,ps at 48\,Gb/s); the shaded band
        above it is the infeasible region. The vertical dashed line at
        30\,mm is the calibration reach, and the gold star on the energy
        panels marks the corresponding measured operating point from the
        ISSCC\,2026 UCIe-compliant 48\,Gb/s silicon prototype~\cite{mondal2026ucie}.
    }
    \label{fig:energy_delay_vs_reach_16L}
\end{figure*}

\subsection{Cross-Node Reach Sweep at 48\,Gb/s}
\label{sec:reach_sweep_16L}

We stress the generator across technology and physical-channel axes on a
16-lane UCIe Standard link (48\,Gb/s/lane, 16-UI $\approx$333\,ps budget)
on three process nodes: TSMC 16\,nm (0.8\,V), TSMC 65\,nm (1.0\,V), and
GF 45\,nm. Reach is swept from 2 to 50\,mm (2.5\,mm grid up to 30\,mm,
calibrated to the ISSCC 2026 prototype~\cite{mondal2026ucie}), and 5\,mm beyond. For every (node, reach) point the full
flow from Section~\ref{sec:coopt} is executed and the latency budget is
applied as a hard feasibility filter. We report three frontier
selections: \emph{balanced} (knee), \emph{best-power}, and
\emph{best-delay}.

Figure~\ref{fig:energy_delay_vs_reach_16L} shows that the three nodes
separate cleanly. TSMC 16\,nm delivers the lowest energy per bit
($\sim$0.77\,pJ/bit at 30\,mm vs.\ 1.2--1.4\,pJ/bit elsewhere) because
its $0.8$\,V supply gives a quadratic switching-energy advantage.
GF 45\,nm produces the lowest delay at every reach ($\sim$60\,ps at
30\,mm for best-delay vs.\ $\sim$125\,ps for TSMC 65\,nm) thanks to
faster devices, but at higher energy than TSMC 65\,nm---it is the
\emph{fast} node rather than the \emph{efficient} one.

The curve shapes also expose how the selector trades the two metrics.
Best-power energy is nearly linear in reach because the selector spends
latency slack freely up to the 333\,ps ceiling; the corresponding delay
trace (panel~e) is therefore jagged. The best-delay column is the
mirror image: smooth delay (panel~f) and noisier energy (panel~c).
A step-up in delay near 30--35\,mm appears across all selections: the
organic channel is RC-limited and its Elmore constant grows
super-linearly, so past $\sim$30\,mm the co-optimizer rebalances the
driver (stage count is integer, producing a step rather than a ramp).
E.g., at 35\,mm the TSMC 16\,nm best-power design doubles its stage
count from 6 to 12; past this transition, the curves resume a smooth
reach dependence.


\section{Conclusion}
\label{sec:conclusion}

We presented \name{}, an automated framework that embeds a distributed $\pi$-ladder channel RC model directly into SPICE characterization, producing Liberty timing arcs with intrinsic channel propagation delay. This bridges the gap between high-level approximations and detailed EM models, enabling rapid design-space exploration without packaging expertise.

The framework automates the full pipeline from JSON configuration to Liberty/Verilog/LEF collateral, with automated termination/equalization selection and TX/RX co-optimization. Multi-PDK support (TSMC 16/65\,nm, GF 45\,nm) enables exploration across technology and packaging dimensions in minutes.

Key results show that package-level decisions dramatically impact PPA: organic substrates unexpectedly outperform silicon interposers at longer reaches ($\sim$10\,mm+). Future work extends to higher data rates where inductance matters and adds 3D chiplet support.

\bibliographystyle{ACM-Reference-Format}
\bibliography{references}

\end{document}